\documentclass[12pt]{iopart}

\usepackage{iopams}

\expandafter\let\csname equation*\endcsname\relax

\expandafter\let\csname endequation*\endcsname\relax

\usepackage{amsmath}
\usepackage{amssymb}

\usepackage{amsfonts}
\usepackage{graphicx}
\usepackage{mathrsfs}
\usepackage{dcolumn}
\usepackage{bm}
\usepackage{mathbbol}
\usepackage{color}
\newcommand{\ket}[1]{\left\vert#1\right\rangle}

\newcommand{\bra}[1]{\left\langle#1\right\vert}

\begin{document}

\def\tr{\text{tr}}
\def\Tr{\mbox{Tr}}

\newtheorem{lem}{Lemma}

\newtheorem{theo}{Theorem}

\title{Increase of quantum volume entropy in presence of degenerate eigenenergies}
\author{Michele Campisi}
\address{NEST, Scuola Normale Superiore \& Istituto Nanoscienze-CNR, I-56126 Pisa, Italy}
\date{\today }

\begin{abstract}
The entropy of a classical thermally isolated Hamiltonian system is given by the logarithm of the measure of phase space enclosed by the constant energy hyper-surface, also known as volume entropy. It has been shown that on average the latter cannot decrease if the initial state is sampled from a classical passive distribution. Quantum extension of this result has been shown, but only for systems with a non-degenerate energy spectrum. Here we further extend to the case of possible degeneracies.
 \end{abstract}

\maketitle

\section{Introduction}
Since the birth of statistical mechanics, it has been recognised that the logarithm of the volume of phase space enclosed by the constant energy surface of a classical thermally isolated systems (the so called volume entropy), is a good expression for its thermodynamic entropy. The mathematical and physical foundation of the volume entropy have been addressed various times in the history of statistical physics \cite{Helmholtz95INBOOK,Boltzmann84CJ98,GibbsBook,Hertz10AP338a,Einstein11AP34,Schlueter48ZNA3,Munster69book,Berdichevsky97Book,Pearson85PRA32,Adib04JSP117,Campisi05SHPMP36,Dunkel06PHYSA370,Campisi10AJP78} and is currently the object of further investigation \cite{Dunkel14NATPHYS10,Hilbert14PRE90,Campisi15PRE91,Hanggi16PTRSA2064}.

One property of the volume entropy that matches the corresponding property of thermodynamic entropy is that, when the adiabatic theorem holds, it remains constant in time \cite{Hertz10AP338a,Berdichevsky97Book}. The question then naturally arises of what happens to the volume entropy when the adiabatic theorem does not hold, e.g., if the timescale of variation of $\lambda$ is not slow compared to the timescales in the system. Is it increasing as one expects for the thermodynamic entropy? A positive answer to this question has been given in \cite{Campisi08SHPMP39} in a statistical sense. If the initial state is sampled randomly from a passive distribution [i.e., a distribution of the form $\rho(\mathbf{z})= f(H_0(\mathbf{z}))$, with $f$ a monotonic decreasing function, and $H_0(\mathbf{z})$ the Hamiltonian at initial time], then the statistical expectation of the volume entropy at a time $t>0$ is larger than its initial expectation value. 

A second question that arises regards how to extend the notion of volume entropy  to the quantum case. The answer is not unique and at least two proposals exist in the literature. One proposal defines the quantum counterpart of the phase space volume as $\hat \Omega(E) = \Tr\, \Theta [E-\hat H]$ \cite{Dunkel14NATPHYS10}. The other defines a ``number operator'' as that operator whose eigenvectors are the Hamiltonian eigenvectors $|\psi_k\rangle$ and whose eigenvalues are the corresponding labels $k$, that order the the eigen-energies in increasing fashion $\varepsilon_1 < \varepsilon_2 <\dots< \varepsilon_N$ . The ``volume entropy operator'' is then the logarithm of the ``number operator''\cite{Campisi08SHPMP39,Tasaki00arXivb,Campisi08PRE78b}. An advantage of the latter definition is that it allows for a generalisation of the classical results mentioned above. A drawback is that it applies to systems with non-degenerate spectra only, thus excluding interesting physical scenarios, e.g. the joining of two identical systems \cite{Joshi13EPJB86}, or their disjoining. The purpose of this work is accordingly to extend the notion of ``volume entropy operator''  to possibly degenerate spectra and to investigate its behaviour under a generic, non-necessarily adiabatic, quantum evolution.
\section{Quantum volume entropy in presence of degenerate eigenenergies}
We consider the entropy associated to a thermally isolated classical Hamiltonian system, with  a parameter ($\lambda$) dependent Hamiltonian $H(\mathbf{z},\lambda)$  in the expression given by Gibbs \cite{GibbsBook}:
\begin{align}
S(E,\lambda) \doteq \ln \Omega(E,\lambda)
\label{eq:S(E,t)}
\end{align}
where $\Omega(E,\lambda)=\int_{V(E,\lambda)} d\Gamma $ denotes the measure of the portion $V(E,\lambda)$  of phase space $\Gamma=\mathbb R^{2f}$ which at fixed parameter $\lambda$  has energy below $E$: $V(E,\lambda)\doteq \{\mathbf{z} \in \Gamma | H(\mathbf{z},\lambda)\leq E\}$, \cite{Khinchin49Book}. As mentioned it is an adiabatic invariant \cite{Hertz10AP338a,Berdichevsky97Book}, and when the adiabatic theorem is not satisfies, on average it cannot decrease if the initial state is sampled randomly from a classical passive distribution [i.e., a distribution of the form $\rho(\mathbf{z})= f(H(\mathbf{z},\lambda_0))$, with $f$ a monotonic decreasing function, here $\lambda_0$ is the value taken by $\lambda$ at time $t=0$].

An analogous result has been proven for quantum mechanical systems with non-degenerate spectrum in \cite{Campisi08SHPMP39,Tasaki00arXivb}. The quantum mechanical treatment requires the introduction of a quantum mechanical counterpart of the classical volume $\Omega(E,\lambda)$. One possible choice is to consider the operator \cite{Campisi08SHPMP39,Tasaki00arXivb,Campisi08PRE78b}
\begin{align}
\hat \Omega(\lambda) = \sum_{k=1}^N k | \psi_k^\lambda \rangle\langle \psi_k^\lambda|
\end{align}
where $N$ is the dimension of the Hilbert space, $| \psi_k^\lambda \rangle$ denote the eigenvectors of the $\lambda$-dependent Hamilton operator
\begin{align}
\hat H(\lambda) | \psi_k^\lambda \rangle =\varepsilon_k(\lambda)  | \psi_k^\lambda \rangle
\end{align}
The spectrum is assumed to be non degenerate for all $\lambda$'s, and the eigenvalues are ordered in an increasing fashion:
\begin{align}
\varepsilon_k(\lambda)>\varepsilon_m(\lambda),\quad  {\rm for }\, \, \,   k>m \quad (m,k \in \{1,2, \dots N\}) 
\end{align}
Accordingly, the expectation of the ``number operator'' $\hat \Omega(\lambda)$ on an energy eigenstate  $| \psi_k^\lambda \rangle$ gives the ordering label $k\in \{1,2, \dots N\}$, saying how many eigenstates exist with energy not above $\varepsilon_k(\lambda) $. This is in analogy with the quantity  $\Omega(E,\lambda)/h^f$ roughly saying how many shells of some unit measure $h^{f}$ exist at some fixed $\lambda$ below the shell of energy $E$. One defines accordingly  the entropy operator as follows
\begin{align}
\hat S(\lambda) \doteq  \ln \hat \Omega(\lambda) = \sum_{k=1}^N \ln  k | \psi_k^\lambda \rangle\langle \psi_k^\lambda|
\label{eq:S-non-deg}
\end{align}
Consider now the case when $\lambda$ changes in time according to a time dependent protocol
\begin{align}
\lambda :  [0,\tau] & \rightarrow  \mathbb R \nonumber \\
t & \mapsto \lambda_t
\end{align}
The time dependent Hamiltonian $\hat H(\lambda_t)$ generates a unitary evolution $U_t$ according to the Schr\"odinger equation
\begin{align}
i\hbar \frac{\partial}{\partial t} U_t = \hat H(\lambda_t) U_t \, ,\qquad U_0= \mathbb 1
\end{align}
Consider an initial state $\rho(0)$ and its evolved $\rho(t)=U_t \rho_0 U_t^\dagger$, and consider the expectation of the volume entropy operator at time $t$
\begin{align}
S(t)\doteq \rho(t) \hat S(\lambda_t)
\end{align}
 The following has been proven
 \cite{Campisi08SHPMP39,Tasaki00arXivb}:
\begin{theo}\label{theo:theo1}
Let $ \varepsilon_k(\lambda_t) > \varepsilon_m(\lambda_t)$  for $k>m$ and all times $t \in [0,\tau]$.
If there exist a probability distribution $\{p_k\}, k \in \{1,2,\dots N\}$ such that $ p_k \leq p_m$  for $k>m$ and
$
\rho(0) = \sum_{k=1}^N p_k  | \psi_k^{\lambda_0} \rangle\langle \psi_k^{\lambda_0}|
$
then
 \begin{align}
S(\tau) \geq S(0)
\label{eq:theo1}
\end{align} 
\end{theo}

The proof is based on the following \cite{Allahverdyan02PHYSA305}
\begin{lem}\label{lem:lem1}
Let $\{p_k\},\{p'_k\}$, $k=1\dots N$,  be two probability distributions. Let $p_k\leq p_m$ for $k>m$.
Let $f_k$ be a non-decreasing real sequence. If there exists a doubly stochastic matrix $\mathbf{A}$ with elements $A_{k,m}$ such that  $p'_k= \sum_{m=1}^N A_{k,m}p_m$, then 
\begin{align}
\sum f_k p'_k \geq \sum f_k p_k
\end{align}
\end{lem}

The thesis of Theorem \ref{theo:theo1} follows by writing  $S(\tau)=\sum_k p'_k \ln k$ and $ S(0)=\sum_k p_k \ln k$ and by noticing that the population $p_m'$ of state $|\psi_m^{\lambda_\tau}\rangle$ at time $\tau$ is linked to the initial population $p_k$  by the expression $p'_k= \sum_n p(k|n) p_n$ where the quantum mechanical transition probabilities 
$p(m|n)= |\langle \psi_m^{\lambda_\tau}|U_t| \psi_n^{\lambda_0}\rangle|^2$ form a doubly stochastic matrix.

Note that the non-degeneracy condition $\varepsilon_k(\lambda_t)>\varepsilon_m(\lambda_t)$ implies that no level crossing occurs during the dynamics. This implies that under the further condition of a slow driving the quantum adiabatic theorem \cite{Messiah62Book} applies. In that case the inequality  (\ref{eq:theo1}) turns into an equality.

In case the spectrum has possibly some degeneracies at a given $\lambda$
we order the eigenenergies in a non-decreasing fashion
\begin{align}
\varepsilon_k(\lambda)\geq \varepsilon_m(\lambda), \forall\,  k>m\in \mathbb \{1,2, \dots N\}
\end{align}
Note the $\geq$ sign, accounting for the possibility that two or more states have the same energy. Let 
\begin{align}
A_k(\lambda)=\{q\in \mathbb \{1,2, \dots N\} | \varepsilon_q(\lambda)=\varepsilon_k(\lambda)\}
\end{align}
be the set of all indices such that for a given $\lambda$ the corresponding energy is equal to $\varepsilon_k(\lambda)$. Let $g_k(\lambda)={\rm card}(A_k(\lambda))$, be its cardinality, namely the degree of degeneracy of the energy eigenvalue $\varepsilon_k(\lambda)$. Let $N_k(\lambda)=\rm{card}\{q \in \mathbb \{1,2, \dots N\} | \varepsilon_q(\lambda)\leq \varepsilon_k(\lambda)\}$ be accordingly the number of states with energy not above $\varepsilon_k(\lambda)$. In analogy with the non-degenerate case one can define the ``number operator'' as $\hat \Omega(\lambda) = \sum_kN_k(\lambda)  | \psi_k^\lambda \rangle\langle \psi_k^\lambda| $ and the entropy operator as $\hat S(\lambda) = \sum_k\ln N_k(\lambda)  | \psi_k^\lambda \rangle\langle \psi_k^\lambda| $. From the physical point of view the latter definition would however present a problem. Imagine the system is at equilibrium in a statistical mixture $\rho= \sum p_k  | \psi_k^\lambda \rangle\langle \psi_k^\lambda|$ at some fixed $\lambda$ where same energy states are equally populated: $p_k=p_m$ if $k,m \in A_k(\lambda)$. Imagine all states are non-degenerate apart from states $| \psi_k^\lambda \rangle, | \psi_{k+1}^\lambda \rangle$, which are doubly degenerate. It is $N_k(\lambda)=N_{k+1}(\lambda)=k+1$, and $p_k=p_{k+1}$. Imagine now that an infinitesimal perturbation $\epsilon\hat V$  lifts the degeneracy, so that we have the new numbers $N^\epsilon_k(\lambda)=k$, $N^\epsilon_{k+1}(\lambda)=k+1$. While  the populations $p_k$ and $p_{k+1}$ would be affected at most by terms of $O(\epsilon)$, the expectation of $\hat S(\lambda)$ would change by the finite amount $\Delta S=p_k[\ln k -\ln (k+1)]+O(\epsilon)$. We regard such finite jump associated to an arbitrarily small perturbation as unphysical. A definition of ``volume entropy operator'' that does not suffer from the degeneracy-lift issue is the following:
\begin{align}
\hat S(\lambda) \doteq \sum_{k=1}^N \ln \mathcal N_k(\lambda) | \psi_k^\lambda \rangle\langle \psi_k^\lambda|
\label{eq:S-deg}
\end{align}
where 
\begin{align}
\mathcal N_k(\lambda) \doteq\left[\prod_{q \in A_k(\lambda)} q\right]^{\frac{1}{g_k(\lambda)}}
\label{eq:N-deg}
\end{align}
is the geometric average of $q$ over the set $A_k(\lambda)$. Accordingly $
\ln \mathcal N_k(\lambda) = g_k^{-1}(\lambda)\sum_{q \in A_k(\lambda)} \ln q
$
is the arithmetic average of $\ln q$ over the set $A_k(\lambda)$. Note that in the case of a non-degenerate spectrum it is $\mathcal N_k(\lambda)=k$ and one recovers the definition in Eq. (\ref{eq:S-non-deg}).
With the definitions in (\ref{eq:S-deg},\ref{eq:N-deg}) we can state the following

\begin{theo}\label{theo:theo2}
Let $ \varepsilon_k(\lambda_t) \geq \varepsilon_m(\lambda_t)$  for $k>m$ and all times $t \in [0,\tau]$.
If there exist a probability distribution $\{p_k\}, k \in \{1,2,\dots N\}$ such that $ p_k = p_m \forall k,m \in A_k(\lambda_0)$, $p_k \leq p_m$  for $k>m$ and
$
\rho(0) = \sum_{k=1}^N p_k  | \psi_k^{\lambda_0} \rangle\langle \psi_k^{\lambda_0}|
$
then
 \begin{align}
S(\tau) \geq S(0)
\label{eq:theo2}
\end{align} 
\end{theo}

\textbf{Proof.} 
We first prove the thesis by assuming that at time $t=\tau$ the spectrum is non-degenerate, i.e. $\mathcal N_k(\lambda_\tau)=k$. It is $S(0)= \sum p_k \ln \mathcal N_k(\lambda_0)$.
Due to the assumption $p_k=p_m \forall k,m \in A_k(\lambda_0)$ it is $S(0)=\sum_k p_k \ln k$. 
On the other hand $S(\tau)= \Tr \rho(\tau) \hat S(\tau)= \sum p'_k \ln \mathcal N_k(\tau)= \sum p'_k \ln k$. Since $p'_k= \sum_n p(k|n) p_n$ with $p(k|n)$ forming a  doubly stochastic matrix the thesis follows from Lemma \ref{lem:lem1}.

Let us now consider the case when the spectrum at time $t=\tau$ possibly have degeneracies. Let us imagine for illustrative proposes that all states are non-degenerate apart from states $| \psi_k^{\lambda_\tau} \rangle, | \psi_{k+1}^{\lambda_\tau} \rangle$, which are doubly degenerate. Their contribution to $S(\tau)$ is $p'_k\ln \mathcal N_k(\lambda_\tau)+p'_{k+1}\ln \mathcal N_{k+1}(\lambda_\tau)=(1/2)(p'_k+p'_{k+1})[\ln k+\ln (k+1)]$. We can re-express that as $\bar{p}_k \ln k + \bar{p}_{k+1} \ln (k+1)$, where $\bar{p}_k=\bar{p}_{k+1}=(1/2)(p'_k+p'_{k+1})$. More generally, in case of many degenerate subspaces of arbitrary dimension, by introducing the new probabilities
\begin{align}
\bar{p}_k = \frac{\sum_{q\in A_k(\lambda_\tau)} p'_q}{g_k(\lambda_\tau)}
\end{align}
the final entropy expectation reads $S(\tau)=  \sum p'_k \ln \mathcal N_k(\lambda_\tau)= \sum \bar{p}_k \ln k$. It now remains to be demonstrated that the $\bar{p}_k$ are linked to the $p_k$ by a doubly stochastic matrix.

Using vector notation $\mathbf{p}'=(p'_1, p'_2, \dots p'_N)^{T}$, $\bar{\mathbf{p}}=(\bar{p}_1, \bar{p}_2, \dots \bar{p}_N)^T$. It is $\bar{\mathbf{p}} = \mathbf V \cdot \mathbf{p}'$ where $\mathbf{V}$ is a block diagonal matrix whose blocks are of the form 
\begin{align}
\frac{1}{g}\left(\begin{array}{cccc}1 & 1 & \cdots & 1 \\1 & 1 & \cdots & 1 \\\vdots & \vdots & \ddots & \vdots \\1 & 1 & \cdots & 1\end{array}\right)
\end{align}
with $g$ the dimension of the block. In $\mathbf{V}$ the various blocks have the dimension of the corresponding degenerate subspaces. Each block is doubly stochastic (the sum of the elements in any row or column is $1$), 
and so is the matrix $\mathbf{V}$ itself. Let $\mathbf{P}$ be the matrix whose elements are the transition probabilities $p(m|n)$ and let $\mathbf{p}=(p_1,p_2, \dots p_N)^T$. It is $\mathbf{p}'=\mathbf{P}\cdot \mathbf{p}$. Then $\bar{\mathbf{p}}=\mathbf{V}\cdot \mathbf{p}'=(\mathbf{V}\cdot \mathbf{P}) \cdot \mathbf{p}$.
Since both $\mathbf{V}$ and $\mathbf{P}$ are doubly stochastic, so is their product $\bar{\mathbf{P}}=\mathbf{V}\cdot \mathbf{P}$ that is 
\begin{align}
\sum \bar{p}_k = \sum_n \bar{p}(k|n) p_n
\end{align}
where $\bar{p}(k|n)$ are the elements of the doubly stochastic matrix $\bar{\mathbf{P}}$. The thesis then follows from  Lemma \ref{lem:lem1}.

\section{Example}
As an example we consider the sudden breaking of a XX spin chain of length $N$ into two identical XX spin chains of length $N/2$. Studies regarding the joining of such spin chains were reported in  \cite{Joshi13EPJB86,Apollaro15PSCT165}. 

The pre-breaking Hamiltonian reads
\begin{align}
H_i &= \frac{h}{2}\sum_{j=1}^{N} {\sigma_j ^z} - \frac{J}{4}\sum_{j=1} ^{N-1} {[\sigma_j ^x \sigma_{j+1} ^x + \sigma_j ^y \sigma_{j+1} ^y]}\, .
\end{align}
where $\sigma_j^\alpha$, $j=1 \dots N$, $\alpha=x,y,z$, denoting the Pauli matrices of  the $j$-th spin.
The post-breaking Hamiltonian reads:
\begin{equation}
H_f = H_A + H_B \,
\end{equation}
where
\begin{align}
H_A &= \frac{h}{2}\sum_{j=1}^{N/2} {\sigma_j ^z} - \frac{J}{4}\sum_{j=1} ^{N/2-1} {[\sigma_j ^x \sigma_{j+1} ^x + \sigma_j ^y \sigma_{j+1} ^y]}\, \\
H_B &= \frac{h}{2}\sum_{j=N/2+1}^{N} {\sigma_j ^z} - \frac{J}{4}\sum_{j=N/2+1} ^{N-1} {[\sigma_j ^x \sigma_{j+1} ^x + \sigma_j ^y \sigma_{j+1} ^y]}\,
\end{align}

\begin{figure}[t]
\includegraphics[width=\linewidth]{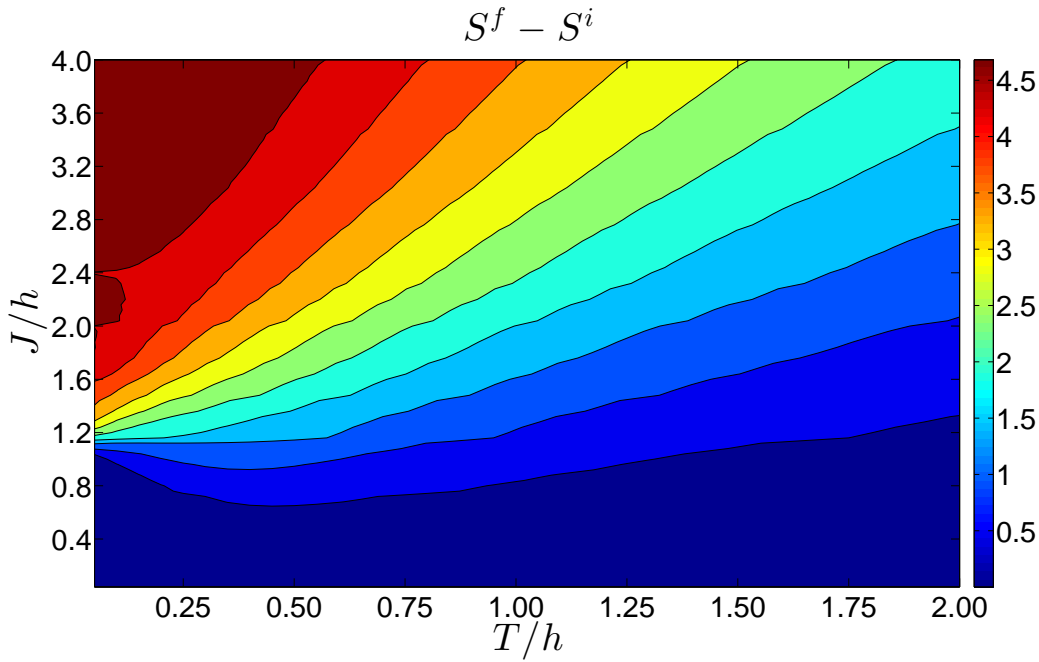}
\caption{Change in the expectation of volume entropy operator for the sudden beak-up of a XX chain as a function of interaction strength $J$ and temperature $T=1/\beta$.}
\label{fig:1}
\end{figure}

The Hamiltonians $H_A,H_B,H_f$ all represent XX spin chains of different lengths $L$. 
The spectrum of the full chain is given by
\cite{Lieb61AP16,Mikeska77ZPB26}
\begin{align}
E^i_\mathbf{n}=&\sum_{k=1}^{N} \varepsilon_{k}(N) n_k
\end{align}
where $\mathbf{n}=(n_1, n_2, \dots, n_N)$, denotes the occupation of each fermionic mode, $n_i=0,1$ and 
\begin{align}
\varepsilon_{k}(N)&=h+ J\cos\frac{k \pi}{N+1}
\end{align}
is the corresponding energy. The final spectrum is accordingly given by 
\begin{align}
E^f_{\mathbf{m}}=&\sum_{k=1}^{N/2} \varepsilon_{k}(N/2)(m_k +m_{k+N/2})
\end{align}
where $\mathbf{m}=(m_1, m_2, \dots, m_N)$, $m_i=0,1$ with $(m_1, m_2, \dots, m_{N/2})$ denoting the occupation of each fermionic mode in the left chain, and $(m_{1+N/2}, m_{2+N/2}, \dots, m_{N})$ denoting the occupation of each fermionic mode in the right chain. Besides possible accidental degeneracies, the final spectrum has further systematic degeneracies due to the $A\leftrightarrow B$ exchange symmetry. 
We assume an initial Gibbs preparation $\rho_0= e^{-\beta H_i}/Z_i$, with $Z_i = \Tr e^{-\beta H_i} $, that is:
\begin{align}
\rho(0) = \sum_{\mathbf{n}\in  \{0,1\}^{N}} p_{\mathbf{n}}^i \ket{\psi^i_\mathbf{n}}\bra{\psi^i_\mathbf{n}}
\, ,\quad p^i_{\mathbf{n}}=\frac{e^{-\beta E^i_{\mathbf{n}}}} {\sum_{\mathbf{n}\in  \{0,1\}^{N}} e^{-\beta E^i_{\mathbf{n}}}}
\end{align}
where $\ket{\psi^i_\mathbf{n}}$ denotes the full chain eigenvectors.
Ordering the initial eigenvalues $E^i_{\mathbf{n}}$ in increasing fashion, amounts to establish a bijective application
\begin{align}
N^i: \{0,1\}^{N}&\rightarrow \{1,2,3, \dots 2^N\}\\
{\mathbf{n}} &\mapsto N^i_{\mathbf{n}}
\end{align}
such that $N^i_{\mathbf{n}'} > N^i_{\mathbf{n}}$ if $E^i_{\mathbf{n}'} \geq E^i_{\mathbf{n}}$. Such application is not unique because the spectrum has degeneracies. 
 We define the sets
\begin{align}
A^i_{\mathbf{n}} = \{\mathbf{n}' \in \{0,1\}^{N} | E^i_{\mathbf{n}'}= E^i_{\mathbf{n}} \}
\end{align}
of all indexes corresponding to the energy $E^i_{\mathbf{n}}$. The pre-quench volume entropy operator reads accordingly
\begin{align}
\hat S^i = \sum_{\mathbf{n}\in  \{0,1\}^{N}}   \ln \mathcal N^i_{\mathbf{n}} \ket{\psi^i_\mathbf{n}}\bra{\psi^i_\mathbf{n}}
\end{align}
where 
\begin{align}
\ln \mathcal N_{\mathbf{n}}^{i} =\frac{ \sum_ {\mathbf{n}' \in A^i_{\mathbf{n}}} \ln N^i_{\mathbf{n}'}} {{\rm card}(A^i_{\mathbf{n}})}
\end{align}
The pre-quench expectation of volume entropy operator is therefore
\begin{align}
S^i = \sum_{\mathbf{n}\in  \{0,1\}^{N}}  p^i_\mathbf{n} \ln \mathcal N^i_{\mathbf{n}}
\end{align}
Similarly for the post-quench quantities with the symbol $i$ replaced by $f$, and\begin{align}
p^f_\mathbf{m}= \sum_\mathbf{n}  |\langle \psi^f_\mathbf{m}|\psi^i_\mathbf{n}\rangle |^2 p^i_\mathbf{n}
\end{align} 
with $|\psi^f_\mathbf{m}\rangle$ the final eignevectors.
The explicit expression of the quantum mechanical transition probabilities $|\langle \psi^f_\mathbf{m}|\psi^i_\mathbf{n}\rangle |^2 $ is given in the appendix of Ref. \cite{Apollaro15PSCT165}. With them we have computed and plotted the change in the expectation of the volume entropy operator as a function of $T=1/\beta$ and $h$, for $N=10$ and fixed $J$. This corresponds to a $1024 \times 1024$ matrix of transition probabilities. The results are reported in Fig. \ref{fig:1}. The computed values of $S^f-S^i$ are all non-negative in accordance with Theorem \ref{theo:theo2}.

\section{Conclusion}
We have extended the results of Ref. \cite{Campisi08SHPMP39} to the case of possibly degenerate spectra, and have illustrated them using the break-up of a spin-chain as an example.

\section*{Ackowledgements}
This research was supported by a Marie Curie Intra European Fellowship within the 7th European Community Framework Programme through the project NeQuFlux grant n. 623085 and by the COST action MP1209 ``Thermodynamics in the quantum regime''. 

\section*{References}

\providecommand{\newblock}{}

\end{document}